\documentclass[conference]{IEEEtran}
\usepackage{upgreek}

\usepackage{url}

\usepackage{amsmath}

\usepackage[T1]{fontenc}

\usepackage{array}

\usepackage{booktabs}

\usepackage{stfloats}

\usepackage{graphicx}
\graphicspath{{./pdf/}{./jpeg/}{./images/}}
   \DeclareGraphicsExtensions{.pdf,.jpeg,.jpg,.png}

\begin{document}
%
\title{Towards the Forensic Identification and Investigation of Cloud Hosted Servers through Non-Invasive Wiretaps}

\author{
\IEEEauthorblockN{Hessel Schut\\ \and Mark Scanlon\\ \and Jason Farina\\ \and NhienAn LeKhac}
\IEEEauthorblockA{School of Computer Science \& Informatics,\\
University College Dublin,\\
Belfield, Dublin 4, Ireland\\
Email: hessel.schut@klpd.politie.nl, mark.scanlon@ucd.ie, jason.farina@ucdconnect.ie, an.lekhac@ucd.ie}}


%
%



\author{\IEEEauthorblockN{Hessel Schut\IEEEauthorrefmark{1}, Mark Scanlon\IEEEauthorrefmark{2},
Jason Farina\IEEEauthorrefmark{3} and Nhien-An Le-Khac\IEEEauthorrefmark{4}}
\IEEEauthorblockA{\IEEEauthorrefmark{1}Korps Landelijke Politiediensten,\\ Driebergen, Netherlands.\\
\IEEEauthorrefmark{2}\IEEEauthorrefmark{3}\IEEEauthorrefmark{4}School of Computer Science and Informatics,\\
University College Dublin, Ireland.\\
Email: \IEEEauthorrefmark{1}hessel.schut@klpd.politie.nl, 
\IEEEauthorrefmark{2}mark.scanlon@ucd.ie,
\IEEEauthorrefmark{3}jason.farina@ucdconnect.ie,
\IEEEauthorrefmark{4}an.lekhac@ucd.ie}
}

\maketitle


\maketitle

\begin{abstract}
When conducting modern cybercrime investigations, evidence has often to be gathered from computer systems located at cloud-based data centres of hosting providers. In cases where the investigation cannot rely on the cooperation of the hosting provider, or where documentation is not available, investigators can often find the identification of which distinct server among many is of interest difficult and extremely time consuming. To address the problem of identifying these servers, in this paper a new approach to rapidly and reliably identify these cloud hosting computer systems is presented. In the outlined approach, a handheld device composed of an embedded computer combined with a method of undetectable interception of Ethernet based communications is presented. This device is tested and evaluated, and a discussion is provided on its usefulness in identifying of server of interest to an investigation.
\end{abstract}


%
\IEEEpeerreviewmaketitle

\section{Introduction}
The National High Tech Crime Unit (NHTCU) in a European country conducts investigations to crimes targeted at ICT infrastructure, committed using  new technology or methods  that have the potential to be incapacitating to society or have a high impact. This impact can be quantified in terms of  financial losses, incurred  business continuity or recovery costs and loss of goodwill in the form of public trust or confidence. Today, in its investigations the NHTCU often conducts searches in cloud-based data centres to collect evidence by preserving stored data on hosted computer systems or to conduct wiretaps on these computer systems.

In most cases these computer systems are owned by a trusted hosting company that leases these computer systems to end-users. Under subpoena, the hosting company locates the exact computer of which data is requested and hands this server over to the investigators who will assert some other investigative power such as the creation of a forensic copy of data, placing a wiretap on the system or conducting live data forensics.

However, the assistance of the hosting provider is not always possible. Several times each year the NHTCU is confronted with co-located servers on cloud platforms ( computer systems owned by a third party that are hosted in a data centre). Often these co-located servers are part of a larger infrastructure, where the networking devices such as switches and firewalls, are also owned by the customer of the data centre. An investigator may find themselves in a situation where the hosting provider or data centre itself cannot be trusted to provide accurate information. In these cases the investigator is confronted with multiple servers without any documentation about these servers. This poses a challenge in identifying the computer system of interest to the investigation. 

One approach available to the investigator is the use of the European criminal process law which  provides for the interception of network traffic \cite{wetboek}. Because of the impact this may have on privacy, a warrant for interception can only be given by an investigative judge. Analysis of intercepted traffic to the individual computer systems could reveal identifying information about these systems such as upstream source IP-address and Ethernet MAC address and HTTP headers like \texttt{Server} and \texttt{Host}. So while the interception of network traffic may be possible in many cases, the investigator  needs to consider the impact this activity will have on privacy. This is especially true when the discovery process may require the interception of traffic from computer systems that are unrelated to the investigation. This impact may be considered disproportionate and therefore a less reliable method of identifying servers would be utilised. The argument can be made, that, when performing discovery and, when the objective is the preservation of data, the identification of servers and the traffic metadata, as permitted by criminal process law \cite{wetboek},that needs to be gathered to do so is an integral part of this search. Following this interpretation, it follows  that Article 125n \cite{wetboek2} applies too, which states that all information gathered during a search that is of no interest to the investigation needs to be destroyed and that a written report  of the destruction needs to be compiled. When applied to the problem of identification of servers, this means that an audit trail of all identification attempts needs to be kept, where the investigator can mark identifying information as either relevant, or irrelevant. In the latter case data can be destroyed immediately, but a record of destruction needs to be kept. Preferably no information is displayed to the investigator at all, 
other than a confirmation that a pre-determined identification is recognised.

Recent approaches in literature are concerned with the creation of wiretaps for Ethernet networks by placing devices inline \cite{gomez2003receive}\cite{nikkel2006portable}\cite{ossmann2011throwing}\cite{zharinov2013saint}, or requiring administrative access to configure a switch for port mirroring \cite{hegge2001system}. To place a wiretap in-line with the connection to the observed device, the connection to that device needs to be interrupted. This poses no problems when creating a long-term wiretap for network management or security purposes, e.g., to connect an Intrusion Detection System (IDS) sensor \cite{gomez2003receive}. However, interruption of an Ethernet link can be signalled and logged at either side of the connection. This may pose a problem for law enforcement as this can alert an adversary that the connection has been tampered with. 

Therefore, a new technical solution is sought for to intercept Ethernet traffic for the purpose of identification of computer servers. For this, a method needs to be developed that is undetectable by the operator of the computer system. In most network investigation, most computer systems are identified by the public IP address. The solution must at least allow the investigator to determine the IP-addresses used by a computer system. Extending to other properties of network traffic, the new solution should allow users to add other identifying properties of the network traffic that are more appropriate to identify the computer system.

The rest of this paper is organised as follows: Section~\ref{related} shows the related work of this research on different approaches on wiretaps for Ethernet networks. A new approach for the identification of computer server using temporary wiretap s is presented in Section~\ref{approach}. The software components of our device is outlined in Section~\ref{software}. A discussion is provided on the wiretap attachment in Section~\ref{attachment}. An evaluation of the proposed device and an analysis of its performance is provided in Section~\ref{experimentation}. The conclusion and a discussion on future work is outlined in Section~\ref{conclusion}.


\subsection{Aim and Contribution of this Work}
\label{aim}
The aim of the work presented as part of this paper is to expedite digital investigations in a cloud data centre environment. This can be achieved by focusing the investigation at an early stage to pertinent servers through the identification of suspicious or targeted network traffic.

In this paper, a device implementing a method for undetectable interception of Ethernet network traffic for the purpose of identification of computer systems is presented. This new solution was designed to be fit for use in the field during investigation. As a result, an easy to operate handheld device was engineered. This solution captures as little information as possible and only stores data permanently when it is considered relevant to the investigation by the operator. Alternatively, the operator has the possibility to not display any addresses, other than the confirmation of the appearance an address of interest that has been defined. An audit log of all operator actions for reporting is mandatory and, while real time timestamps are desirable at a minimum relative timestamps of activities and discoveries need to be logged.


\section{Related Work}
\label{related}

In this section, related work on wiretaps for Ethernet networks is discussed. This work can be divided in to main categories: passive and active approaches.
\subsection{Active Approaches}
Switch Port Analyser or SPAN \cite{hegge2001system} is a method of intercepting (wiretapping) Ethernet traffic for purposes such as network monitoring. SPAN is a technique where a switch has one or more ports defined as \texttt{mirror} ports to which monitoring devices can be connected. A network administrator can set up a mirroring policy that identifies types of traffic that is being copied by the switch from the receiving port(s) to one or more mirror ports for analysis. In \cite{hegge2001system} the mirrored traffic is distributed to mirror ports in a round-robin manner, thus distributing traffic over multiple ports to make it more likely that the mirror ports can handle the copied data without dropping traffic.

\cite{nikkel2006portable} proposed a forensics evidence collection device called \texttt{PNFEC}. The proposed device uses an embedded computer with multiple network interfaces and open-source software to collect live network evidence from single hosts. The \texttt{PNFEC} is placed inline between a network node of interest and the rest of the network. After being placed inline between the network node and the network, the device acts as a transparent Ethernet bridge \cite{ieee2010ieee} between the observed network node and the rest of the network. The bridge is created using the \texttt{brconfig} command  in the OpenBSD operating system, thus using the bridge kernel driver to forward traffic from one network interface to another.

An inline wiretap device is proposed in \cite{zharinov2013saint}. The proposed device is connected in parallel with the network cable to the observed device using alligator clips. A faster and more robust way of connecting to the network cable is left for further work. 

The parallel connection through the proposed device is routed through electromechanical switches (relays) for each wire. These relays are in a normal closed state; therefore the original cable can be cut safely when the relays are unpowered. When power is applied to the relays, the relay contacts divert the Ethernet signals to two network adapters that are operating in a transparent bridge, as described in \cite{nikkel2006portable}. Should power to the device fail, the relays fall back to a state where the cable is directly connected again, bypassing the bridge. This setup also makes it possible to conduct Man-in-the-Middle (MitM) attacks and selectively block traffic.

\subsection{Passive Approaches}
Instead of actively bridging Ethernet \cite{nikkel2006portable}\cite{zharinov2013saint} or copying network traffic in switches \cite{dpkt}, it is possible to wiretap 10-Mbit/sec and 100-Mbit/sec Ethernet links passively using modified unshielded twisted pair (UTP) cables \cite{gomez2003receive}. In \cite{gomez2003receive}, four methods of modifying UTP cables to create receive-only UTP cables are evaluated for use in network intrusion detection systems (IDS). The modified cables provide the necessary signals to enable the link on an upstream hub or switch and to introduce errors in data sent from the attached sniffer to prevent an upstream hub or switch from recognising this data. The cables rely on an Ethernet hub to forward packets to the cable, as described in \cite{gomez2003receive}. The proposed cables are compared to Test Access Ports (TAPs) that sit in-line between the hub or switch and the observed device and SPAN ports, as described above. A TAP is described in \cite{gomez2003receive} as: ``a device that allows to examine network traffic without causing any data stream interference'', acting at OSI layer 1, the physical layer. A connection diagram for a TAP is also given in \cite{ieee2010ieee}.

An implementation of a TAP is the Throwing Star LAN Tap, designed by \cite{ossmann2011throwing}. The Throwing Star LAN Tap is a star shaped printed circuit board with four RJ-45 connectors on the points of the star. The circuit board needs to be placed in-line between the observed device and an upstream network device, e.g. a hub or a switch. The Throwing Star LAN Tap inserts two 220pF capacitors in the signal path of the observed device to limit the bandwidth of its connection. This limited bandwidth prevents the target devices to auto-negotiate a link on 1000 Mbits/s, which cannot be wiretapped using the Throwing Star LAN Tap and forces them to revert to a 100 Mbit/s link which can be monitored with the Throwing Star LAN Tap. 

The Throwing Star LAN Tap is published as open-source hardware by its designer and sold commercially as a kit of components to be assembled by the user.
\subsection{Discussion}
To create wiretaps for Ethernet communications the previous work either places devices inline \cite{gomez2003receive}\cite{nikkel2006portable}\cite{ossmann2011throwing}\cite{zharinov2013saint}, or requires administrative access to configure a switch for port mirroring \cite{hegge2001system}.
This problem is addressed in \cite{zharinov2013saint} where electromechanical switches (relays) are used to divert the connection to a transparent bridge before interrupting the physical connection to the observed device. However, placing a bridge inline between the observed device and an upstream component, as proposed in \cite{nikkel2006portable} and \cite{zharinov2013saint} may alter the physical layer characteristics of the Ethernet connection, i.e., the link speed and duplex type of the connection may differ from the original device. Bridging the connection may also introduce additional latency to the network traffic, which may be detectable from either the observed device or connected devices.

The Throwing Star LAN Tap \cite{ossmann2011throwing} is constructed deliberately to alter the characteristics of the Ethernet connection that is to be observed to ascertain that it conforms to a media type that can be wiretapped passively.

Changes in the physical layer characteristics may interrupt connectivity, either temporarily while the link renegotiates to a media type that both sides of the connection can support, or permanently when no common media type can be agreed on given the changed configuration. This may be the case if the network interface at either side does not support auto-negotiation of the physical layer media type, or when this auto-negotiation is administratively overruled \cite{raspbian2015about}. 

Development of a fast and reliable method of attaching a wiretap device to a network cable is stated as a requirement in \cite{zharinov2013saint}, but left open for future research.  

The methods proposed in this paper focus on the identification of computer systems using temporary wiretaps. This problem may be very specific to law enforcement; it is not addressed in the literature found on the subject. The proposed method addresses the open problems found in previous work; i.e. creating wiretaps without interfering with the Ethernet physical layer in a way that is detectable from the observed device(s) and making reliable temporary connections to the network cable(s) that connects to the observed device(s).

\section{New Approach for Cloud Server Forensics Using Temporary Wiretaps}
\label{approach}
\begin{figure}[!t]
\centering
\includegraphics[width=0.4\textwidth]{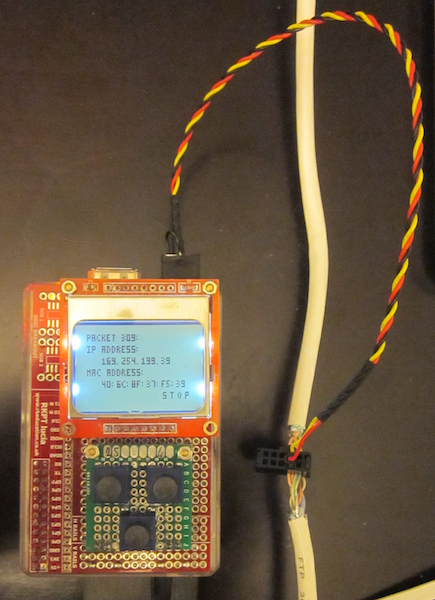}
\caption{The Proposed Device Connected to an Ethernet Cable}
\label{fig:device}
\end{figure}

To address the issues as defined in previous sections, this paper proposes a handheld device that passively intercepts traffic from an Ethernet twisted-pair cable, as can be seen in Figure~\ref{fig:device}. An embedded computer integrated in the device processes the intercepted data and displays identifying characteristics of the traffic on a liquid crystal display (LCD). To attach the device to the Ethernet cable of the computer system that is to be identified, a fast and reliable method of creating a test access port (TAP) to an active cable is suggested, as displayed in Figure~\ref{fig:device}.

\subsection{Selection of Hardware and Software}
\subsubsection{Hardware} To process and display the intercepted network traffic, moderate computing resources are required. In fact, the computing platform needs to provide an Ethernet network interface to capture packets on and needs to have input and output ports (general purpose input and output or GPIO) to implement the user interface. Another important requirement is the capability of supporting a common operating system, e.g., Linux. By using a common operating system, a broad selection of existing software can be used for this approach. This has another advantage that most of the development time can focus on the actual application instead of the underlying software infrastructure, e.g., scheduling, network drivers, packet dissectors. The device is designed to be used on location in a data centre moving between different servers to identify pertinent systems. Therefore, a few hours of autonomy on battery power is preferable, but a less hard requirement than the above as wired power should normally be available. Various low-cost embedded boards that are capable of running the Linux operating system are on the market at the time of writing. The actual selection is not critical and given the same underlying operating system, it can be expected that these boards can be used interchangeably with only minor changes. In the approach outlined as part of this paper, the Raspberry Pi platform \cite{raspberry2015what} was chosen as this hardware meets the aforementioned requirements, is affordable and lightweight and can be easily powered by USB battery packs.

\subsubsection{Software}
\label{subsoftware}

\begin{itemize}
\item Linux is chosen as the operating system for the system; specifically the Raspbian \cite{raspbian2015about} distribution, which is a variant of the well-known Debian Linux distribution, adapted to the Raspbery Pi hardware and optimised to the Pi's ARMv6 CPU. As stated in \cite{raspbian2015about}, on the initial releases of Raspbian, over 35,000 software packages were available for the platform. Specifically to this project, the most common network forensic software, e.g., Wireshark and tcpdump, is available as ready built packages on the Raspbian Linux distribution.

\item The Python programming language is chosen to develop the application integrated in the device, because this language is well supported on the Raspberry Pi. Many modules have been written in the Python programming language to interact with the Raspberry Pi's input/ output (I/O) ports, which means that less development time needs to be spent on implementing the details of interfacing with hardware, e.g., an LCD display for the user interface, etc.

    \item Wireshark and \texttt{tshark}: In \cite{zharinov2013saint} different network analysis software programs are compared for the use in an embedded device. The choice in \cite{zharinov2013saint} is the common program tcpdump, for its low memory requirements. The Raspberry Pi, has 512 MB of RAM available, which is enough to support all the compared network analysis software. The program Wireshark has the advantage over tcpdump that it has a command-line variant, \texttt{tshark} that has a configurable output format. Indeed,\texttt{tshark} can be compiled with an internal script interpreter. This interpreter allows for programs written in the programming language Lua to be run within the \texttt{tshark} or Wireshark programs \cite{lamping2004wireshark}. These features make \texttt{tshark} better suited for this application than tcpdump, which produces human-readable output. Human-readable output is generally harder to parse than a structured (XML or delimited text) output format that \texttt{tshark} offers.

\item \texttt{libpcap}, \texttt{pypcap} and \texttt{dpkt}: Another option is to integrate network capture and analysis into the network identification program. Most network analysis software is based on the \texttt{libpcap} library \cite{jacobsen2005tcpdump}, therefore instead of using an external network analysis program, a language binding to the libpcap library can be used instead. This has the benefit that no external programs need to be called, providing better control over the actions on the network capture interface and the captured data. For the Python programming language a \texttt{libpcap} extension module with the name \texttt{pypcap} has been developed \cite{pypcap}. The \texttt{pypcap} module provides raw captured traffic to the user; another module is needed to dissect the captured data. Dissection is the interpretation of captured network traffic, i.e. the decoding of fields in each of the protocols in a packet. A Python module that is capable of dissecting the data that is captured with \texttt{pypcap} is \texttt{dpkt} \cite{dpkt}. For this application it is needed to dissect packets until the IP-header. For this purpose, the combination of \texttt{pypcap} and \texttt{dpkt} library is sufficient as \texttt{dpkt} handles both the Ethernet and IP-header information. Through the use of these libraries, most functionality needed for the system can be integrated in a single program. Though, conceivably additional dissectors are necessary to decode other identifying protocols.
\end{itemize}
\subsection{System Architecture}
First of all, a new hardware user interface was developed for the proposed system.

\begin{figure}[!t]
\centering
\includegraphics[width=0.4\textwidth]{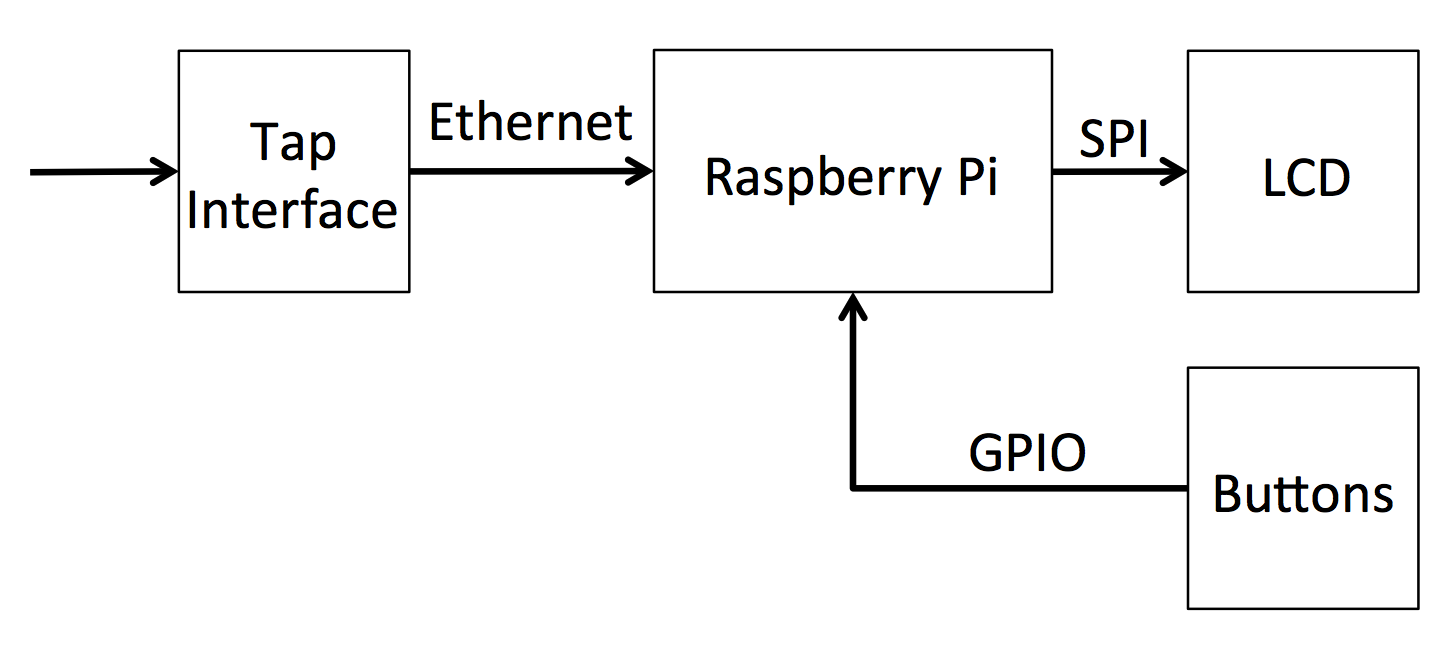}
\caption{System Block Diagram}
\label{fig:system}
\end{figure}

\subsubsection{System Overview} The system is designed to be used as a handheld device and components are selected for their small size. At the time of writing, no enclosure has been selected for the device, though. The hardware of the device consists of four parts:
\begin{itemize}
\item Computer
\item Display
\item Buttons
\item Wiretap Interface
\end{itemize}
The relationship between these parts is shown in the diagram in Figure~\ref{fig:system}.

\subsubsection{Display} The display that is used is a small Liquid Crystal Display (LCD) module based on the PCD8544 48x84 pixels display controller \cite{pcd}. These displays are originally produced for various Nokia mobile telephones (e.g. Nokia 3310 and 5110). The PCD8544 controller is connected to the Raspberry Pi's Serial Peripheral Interface or SPI bus \cite{berger2013raspberry}. The SPI bus is a serial bus that has one master and one or more slaves. Serial transfer is synchronous to a clock that is produced by the master device. In this project the Raspberry Pi acts as the SPI master device. The PCD8544 has no built in character set for displaying text; it is purely a bitmap display. Therefore all text that is to be displayed needs to be generated in software and sent as pixel data to the display. A software module for the Python programming language on the Raspberry Pi is available as open-source software \cite{berger2013raspberry}; this module is extended with alternate fonts for the developed system.

\subsubsection{Buttons} The user interface uses two buttons for user input; these buttons are connected directly to two GPIO lines of the Raspberry Pi as show in Figure~\ref{fig:system}.

These lines are kept on a defined level by means of pull-up resistors, the buttons switch to ground. This means that when a button is pressed the voltage on the corresponding GPIO line will transition from the supply voltage of 3.3 volt to about 0 volt. When the Raspberry Pi observes a falling edge, i.e., the signal on the GPIO line going from the supply voltage to zero volts. This will cause an interrupt that the user-interface software will handle to process user interaction.

\section{Software Components of the Proposed Device}
\label{software}
\begin{figure}[!h]
\centering
\includegraphics[width=0.45\textwidth]{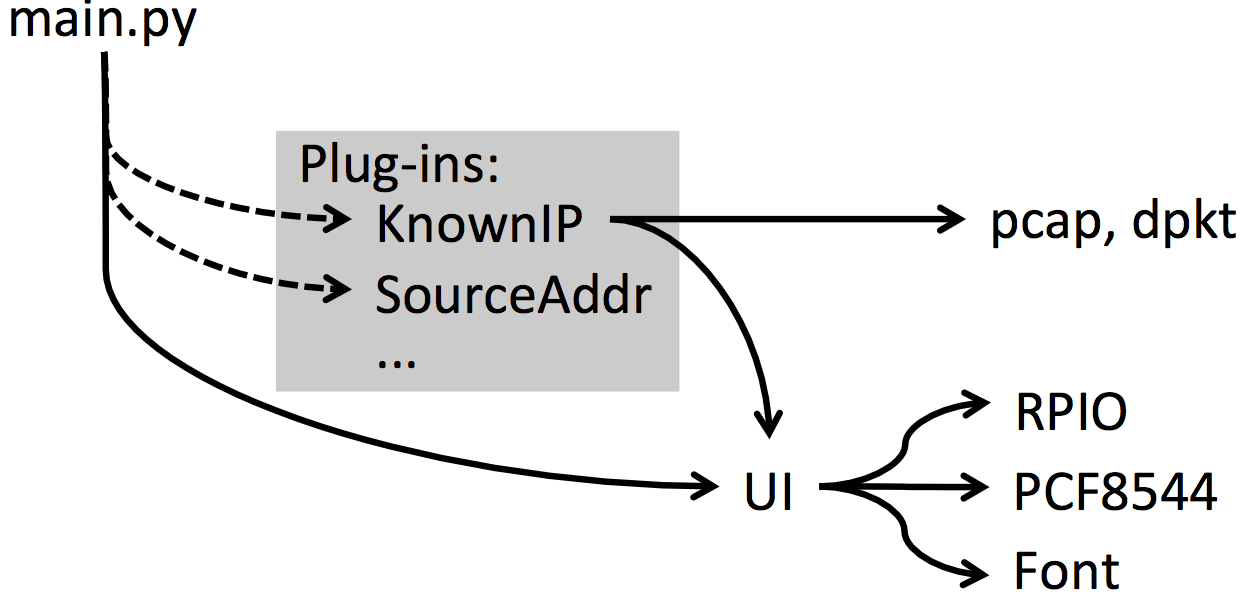}
\caption{Graph Representation of the System's Module Dependencies}
\label{fig:moduledep}
\end{figure}
Most work in our approach went in creating modular software for the proposed device. The software consists of a main application, that dynamically loads and executes plug-in modules. The plug-in modules facilitate augmentation of the system with different means of identifying computer systems. The plug-in modules present the main application lists of parameters that they require or provide. The main application will prompt the user for these parameters and will display them after running a module. This means that plug-in modules do not necessarily need to have a user-interface implemented. All parameters entered and all results provided by the plug-in modules can optionally be logged in an audit log that will aid investigators in reporting their actions. Both the main application and the implemented plug-in modules use a Python module called UI. This UI module implements the user-interface for the system, providing a range of widgets to prompt for data or present data to the user. Two identification plug-in modules have been implemented using this framework; one module \texttt{SourceAddr.py} lists all Ethernet and IP addresses observed on the TAP and presents these addresses. The second module \texttt{KnownIP.py} requires input of a known IP address and will present the user with a counting of packets seen and packets where the source IP address matches the entered known address, confirming it is present in the intercepted traffic. This functionality allows for identification of a computer system without revealing any other information to the investigator. Details of these modules will the described in the following sub-section.
\subsection{Overview}
The architecture described above is depicted in Figure~\ref{fig:moduledep}, as this graph shows, all software necessary for interfacing with the hardware (the modules RPIO, pcf8544 and Font) are abstracted from the other components by the user-interface module UI.
The plug-ins are enumerated by the main application and loaded dynamically when the user selects a plug-in to use, this is denoted by the dashed lines in the graph in Figure~\ref{fig:moduledep}. 
\subsection{User Interface}
The user-interface module UI is responsible for both abstraction of the underlying hardware of the system as well as providing the other components with user-interface widgets to request information from the user or present information to the user. The provided widgets are self-contained to the one who needs no knowledge of the underlying hardware implementation to use the user-interface widgets in other components of the system, e.g., a new plug-in module that is to be created.
\subsection{Main Program}
Using the user-interface widgets from the user-interface module UI, the main program handles the following tasks in order:

\begin{itemize}
\item Initialise the user-interface.
\item Enumerate all plug-in modules.
\item Present a selection list of plug-in modules to the user.
\item Allow for audit log to be kept.
\begin{itemize}
\item When audit log is kept, ask for present date and time.
\end{itemize}
\item Request all parameters that the plug-in module needs from the user, committing these to the audit log, when enabled.
\item Run the plug-in module until it exits, possibly on user request.
\item Present results to the user, committing these to the audit log, when enabled.
\item Rerun the plug-in module with the same parameters on user request.
\item Run another module on user request.
\end{itemize}

Because most of the user-interface handling is abstracted in the UI module, the tasks describe above are to be found in the same order in the main program. To allow for new functionality to be added to the system without alteration of the main program, a plug-in system was designed for this system.

\subsection{Plug-in Modules}
Two plug-in modules have been developed for the application described above. The plug-in modules can be considered the actual solution to the problem as stated in the introduction, the hard- and software described so far supporting these plug-in modules. It is considered necessary, though, because new cases that NHTCU encounters have different requirements, e.g. different identifying properties of the intercepted network traffic should be investigated. It is conceivable that the same framework is not only used for Ethernet but also for wireless devices added to the platform, the plug-in structure allows the platform to be extended for such new use cases. As mentioned before, two plug-in modules have been developed. Both modules aim to solve the problem of finding a host with a known IP address among a number of computer systems connected through Ethernet.

SourceAddr Plug-in Module uses the wiretap technique described in Section~\ref{attachment} to gather upstream network traffic from the computer system where the investigator tries to identify. Of the upstream network traffic both the Ethernet and IP source addresses are decoded and immediately presented to the investigator. When the investigator has received enough identifying information to make a decision, the acquisition and decoding of traffic can be stopped with a button press, all unique source addresses are then handed to the main application. The gathered source addresses can then be committed to the audit log, when this log is enabled. As described in Section~\ref{subsoftware}, the Python modules \texttt{pycap} and \texttt{dpkt} are used for capturing and dissection of network traffic.

KnownIP Plug-in Module has the same goal as the SourceAddr Plug-in Module. However, it identifies the observed host by the source IP address that the host uses upstream. Instead of listing all observed addresses, though, this module only shows a count of packets of which the source IP address matches an IP address that the investigator entered. The purpose of this change is to minimise the impact this tool imposes on the privacy of the users of the computer systems that is used. The SourceAddr Plug-in captures no more than the 34 bytes necessary to parse the Ethernet and IP header of packets to prevent the capture of other data than header information. As IP addresses are considered identifying information, the presentation of this information about computer systems irrelevant to the investigation is undesirable. Therefore, the KnownIP module presents this information to the investigator by only displaying a tally of the number of packets where the source address matches the address of interest of the investigation versus the total number of packets processed.

\begin{figure}[!t]
\centering
\includegraphics[width=0.35\textwidth]{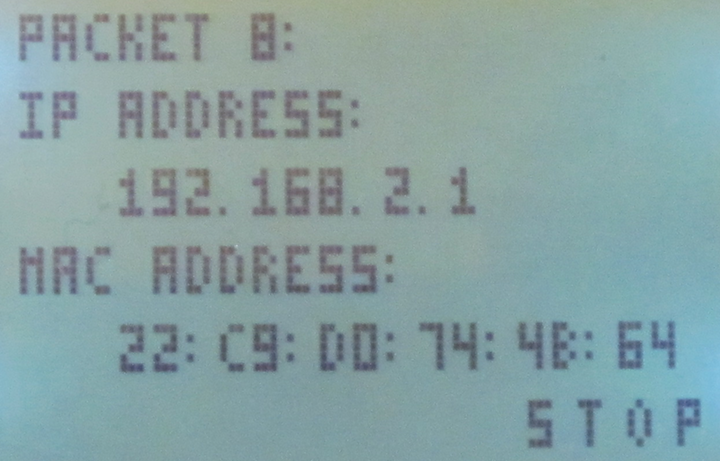}
\caption{User Interface of the \texttt{SourceAddr} Plug-in}
\label{fig:sourceaddr}
\end{figure}

\begin{figure}[!t]
\centering
\includegraphics[width=0.35\textwidth]{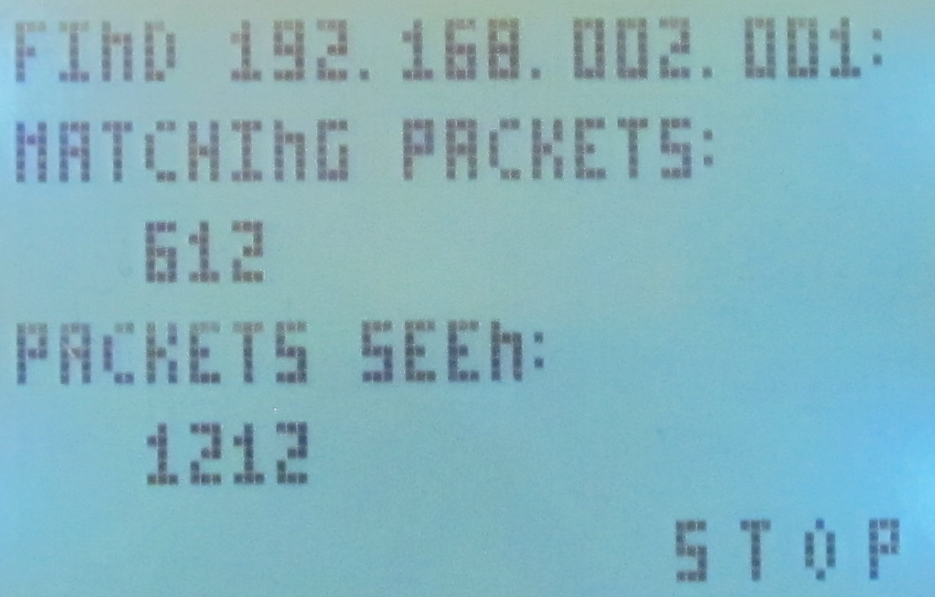}
\caption{Information Presented by the \texttt{KnownIP} Plug-in}
\label{fig:knownip}
\end{figure}

Figure~\ref{fig:knownip} is an example of the information the KnownIP plug-in presents, compared to the user-interface of SourceAddr as shown in Figure~\ref{fig:sourceaddr}. The detail is only the packet counts and it provides no address information that was not known to the investigator in advance. The level of detail is sufficient to identify a server based on matching source addresses, when this match is made, one could consider confirming this match using the detail of the SourceAddr plug-in. As the observed host had transmitted one or more packets of which the source address is matched, it is likely to be of interest to the investigation. This may warrant observing the traffic in more detail to ascertain the match found from the KnownIP plug-in. The SourceAddr plug-in can be run on return to the main application, this choice will be logged in the audit log, when it is enabled. This allows the investigator to write a detailed reasoning behind their actions in a report based on the information provided from the log file.

\subsection{Audit Logging}
Another purpose of the plug-in modules is that audit logging can be enforced through the main application. The reason behind this is that enough abstraction is provided by the framework and that plug-ins can be written by different developers while maintaining a similar usage experience and expectations can be made on a minimum standard of audit logging that will be done. The parameters decisions may have been made upon are deemed the most important elements of an audit log. Together with timestamps in a resolution of at least a minute, these will provide all details the investigator needs to compile a report of their actions. In fact, the Raspberry Pi does not provide a real time clock that keeps accurate time between shutdowns.

In this application, the network interface of the Raspberry Pi is used with a receive-only cable to intercept traffic. Network communication is therefore impossible. This makes it impossible to synchronise the operating systems clock over the network using a time synchronisation protocol (e.g. NTP the Network Time Protocol). A real time clock with battery backed-up hardware can be added to the Raspberry Pi, this will use two more GPIO lines, but was not examined in our approach. Instead, in this approach the user is asked, when logging is enabled, to enter the current date and time. Although unsynchronised, the operating system does update a clock that will be relatively stable considering the time resolution reporting needed and the expected timeframe of the investigators actions. From the date and time entered by the user an offset from the operating system clock is calculated. This offset is then used by the unmodified operating system clock for logging with timestamps of satisfying accuracy. It should also be noted that the choice not to set the operating system clock but to use an offset from this clock has a disadvantage that metadata file will use the unsynchronised operating system time. However, the current plug-in modules do not use any file metadata. The process of entering the current day and time does not take a very long time. When no reporting is necessary, though, it can be experienced as a nuisance therefore the user is given the option to bypass logging. This is a trade-off between enforcing logging and usability. During development the option to bypass logging is appropriate, it can be argued though that in general using, logging should be enforced at all time because the habit of bypassing the logging may arise among users. This habit may subsequently impair the quality of reporting by the investigators.

\section{Wiretap Attachment}
\label{attachment}
As described in Section~\ref{related}, an open problem is the non-invasive attachment of wiretaps to an existing connection. In \cite{zharinov2013saint} alligator clips are proposed as a means of connecting to an Ethernet UTP cable. Alligator clips pierce neither the outer sleeve of the cable, nor the inner insulation of the actual wire strands inside the cable. This means that the insulation of the conductors needs to be removed, which is a delicate and time consuming process. Depending of the mechanical tolerances alligator clips do not offer a reliable connection, the mechanical size of alligator clips poses constraints on the length of wire that can remain twisted, possibly harming the signal integrity of both the intercepted connection and the intercepted traffic. To make quick and reliable connections to the UTP cable leading to the computer system under investigation, a more suitable solution was sought for.

\subsection{Insulation-Displacement Connectors} Insulation-displacement connectors (IDC)are a common type of connector in electronics. These are mostly used on so-called ribbon cables, also called IDC-cables, for the use with IDC technology connectors. More details on IDC can be found in \cite{lee1999insulation}.

\subsection{TAP Creation Using IDCs} When the outer insulating sleeve of a UTP cable is cut lengthwise, the internal wire-pairs can be exposed. Using a suitable tool, in this research a pair of medical artery clamps was used with good results, the wire-pairs in an UTP cable can be kept parallel to each other over a length that is sufficient for crimping an IDC connector over these wire-pairs. The Attached IDC connector now acts as a test access port (TAP), as described in \cite{gomez2003receive}.

\subsection{TAP Attachment to Analysis Device} An Ethernet cable, of which only the receive wire-pair is wired to the RJ45 connector can now be securely connected to the IDC connector by means of a mating connector. In this research a female IDC crimp connector with was used to attach to the UTP cable and a twisted wire pair with an attached two pin header is connected to the IDC connector. The wire pair that is spliced to the connection this way is fed to the receive side of an Ethernet port, this way the TAP can be used for unidirectional interception of Ethernet traffic. Using two Ethernet interfaces for interception would allow for bidirectional interception of the traffic, this is not necessary for the purpose of identification of servers. For the purpose of identification of a host, the source addresses it uses upstream are sufficient.

\section{Experimentation and Results}
\label{experimentation}
In this section, numerous experiments and results with the proposed device are presented.
\subsection{Effects on Transmitted Traffic}
\label{effectstransmitted}
Experimentation was conducted to study the effects of the interaction of the TAP with the traffic flowing over the observed connection. An Ethernet link using 100 Mbit/s full-duplex was established between two hosts running the Mac OS operating system. Using the ping utility, each 10ms a 1,500 byte sized ICMP echo packet was sent from one host to the other. The standard reporting of the ping utility was used to verify successful transmission and reception of all data over the Ethernet connection.

To establish the norm, two runs of 10,000 packets were sent, and the returned ICMP echo replies were counted. This test was repeated with the TAP attached to wire pair 2 and wire pair 3 to measure if any significant influence on the connection could be observed. The results of this experiment are summarised in Table~\ref{tab:effects}. 

\begin{table}[h]
\caption{Effects of the TAP on ICMP Traffic}
\label{tab:effects}
\begin{tabular}{@{}llll@{}}
\toprule
\textbf{Configuration}       & \textbf{Run \#}                               & \textbf{Packets Sent}                                 & \textbf{Packets Received}                             \\ \midrule
TAP not connected            & \begin{tabular}[c]{@{}l@{}}1\\ 2\end{tabular} & \begin{tabular}[c]{@{}l@{}}10,000\\ 10,000\end{tabular} & \begin{tabular}[c]{@{}l@{}}10,000\\ 9,998\end{tabular}  \\ \midrule
TAP connect to wire pair 2   & \begin{tabular}[c]{@{}l@{}}1\\ 2\end{tabular} & \begin{tabular}[c]{@{}l@{}}10,000\\ 10,000\end{tabular} & \begin{tabular}[c]{@{}l@{}}10,000\\ 10,000\end{tabular} \\ \midrule
TAP connected to wire pair 3 & \begin{tabular}[c]{@{}l@{}}1\\ 2\end{tabular} & \begin{tabular}[c]{@{}l@{}}10,000\\ 10,000\end{tabular} & \begin{tabular}[c]{@{}l@{}}9,999\\ 10,000\end{tabular}  \\ \bottomrule
\end{tabular}
\end{table}

The results of these experiments show that in the normal for 20,000 ICMP echo requests sent, 2 ICMP echo response packets were not received. With the TAP connected to wire pair 2, for all 20,000 ICMP echo requests sent, ICMP echo responses were received. With the TAP connected to wire pair 3, for 20,000 ICMP echo requests sent 1 ICMP echo packet was not received. From this, one can conclude that with the TAP connected, the reliability of the connection is not impaired in respect to the normal.

\subsection{Effect of TAP Insertion}
The previous experiment was conducted with a static connection of the tap from the start of each run of the experiment. As the insertion of the receive-only cable form the Raspberry Pi to the TAP may introduce errors that may not arise with a static connection to the TAP, an experiment was conducted using the same experimental setup as used in previous experiment. Instead of using a static connection to the TAP during the run of the experiment, one run was conducted in which during the length of the run the tap was manually switched 102 times in 108 seconds from wire pair 2 to wire pair 3 of the connection where the TAP is connected to. In this run, for 10,000 ICMP echo requests, 9,997 ICMP echo responses were received, the three lost responses do not deviate strongly from the normal that was established in the previous experiment. From this, it can be concluded that the insertion of the receive-only network cable from the Raspberry Pi to the IDC connector used for the TAP does not lead to a significant impairment of the reliably of data transmission over the intercepted Ethernet connection.

\subsection{Effect on Signal Integrity}
An important concern with the attachment of the wiretap to the TAP created using the method described in Section~\ref{attachment} is the impact of this TAP has on signal integrity. During the experiment described in Section~\ref{effectstransmitted} no significant deviations in reliability of data transfer compared to the norm was found.  This, however, does not guarantee correct operation in all environments. Therefore experiments were conducted to measure the behaviour of the proposed method to create a TAP on an active Ethernet connection.

\subsubsection{Signal Attenuation due to the Introduction of a TAP} Because the network interface of the Raspberry Pi may load the signal of the connection to which the TAP is created, attenuation may occur. To measure the level of attenuation of the signal, a cable with at TAP as described in Section~\ref{attachment} was connected between a network interface and an unmanaged Ethernet switch using 100 Mbit/sec Ethernet. The average peak-to-peak signal on the cable was measured using an oscilloscope. An average peak-to-peak voltage level of about 1.42~V was measured. After measuring the average voltage in the normal situation, a Raspberry Pi was connected to the TAP port and the average peak to peak voltage was measured again, resulting in an average peak to peak voltage of 1.20~V, a difference of 220~mV. Expressed in decibels, this is equal to 20$\times$log(1.20~V/1.42~V)  ($\sim$ -1.46dB) attenuation  with a Raspberry Pi attached to the TAP.

\subsubsection{Allowed Attenuation} The Ethernet IEEE standard 802.3-2012 \cite{wg80232012standard} defines the allowable insertion loss for the 100BASE-T media type as: ``The insertion loss of a link segment shall be no more than 14.6 dB at all frequencies between 2 and 16~MHz. This consists of the attenuation of the balanced cabling pairs, connector losses, and reflection losses due to impedance mismatches between the various components of the link segment''. UTP cables are categorised in quality grades, most patch cables used today are category 5 or higher, and for telephony mostly category 3 cabling is used. The Commercial Building Telecommunications Cabling Standard TIA/EIA-568-B.2, Section 4 defines the maximum allowed insertion loss per 100 meter of cable. For category 3 cable, this figure is 13.1dB/100m at 16~MHz. For category 5e cable the maximal allowed insertion loss is 8.2dB/100m. The added attenuation by the TAP may therefore cause problems for a category 3 cable of 100-meter length, although it is unlikely to encounter cables of this length and category in a data centre. So, for a category 5 cable of up to 100 meters in length the TAP will not likely cause any degradation of the signal due to attenuation of the signal on the intercepted network connection. In \cite{cohen2003cabling} measurements of various samples or category 5e UTP cables were taken, the author concludes that many category 5e cables perform better than the TIA/EIA-568-B.2 standard requires. The author also concludes, however, that most significant signal degradation is due to poor connectors, this factor is not calculated in the insertion loss calculations above. 

From the above follows that the introduction of the TAP to an active Ethernet cable does not attenuate the signal on the existing cable significantly, the overall effect may depend on the quality of the cables and connectors used for the intercepted cable. 

\subsubsection{Reflections and Other Signal Distortion} Above, the attenuation that the introduction of a TAP to an active Ethernet imposes on the signal was discussed. Attenuation is not the only effect the TAP may have on the integrity of the transmitted signals. Reflections due to mismatches in impedance are another effect that may cause problems to the signal integrity of the link the TAP is introduced to. An experiment was conducted to find any reflections that may distort data transmitted over the cable the TAP is introduced to. In this experiment, one wire pair of an UTP cable was terminated in a 100$\Omega$ resistor at the far end, at the near end, the cable was connected to a square wave signal generator and an oscilloscope. The sharp rising edge of the square wave is used to measure transmission characteristics of the cable; this method is known as time-domain reflectometry. 

\begin{figure*}[t]
\centering
\includegraphics[width=0.6\textwidth]{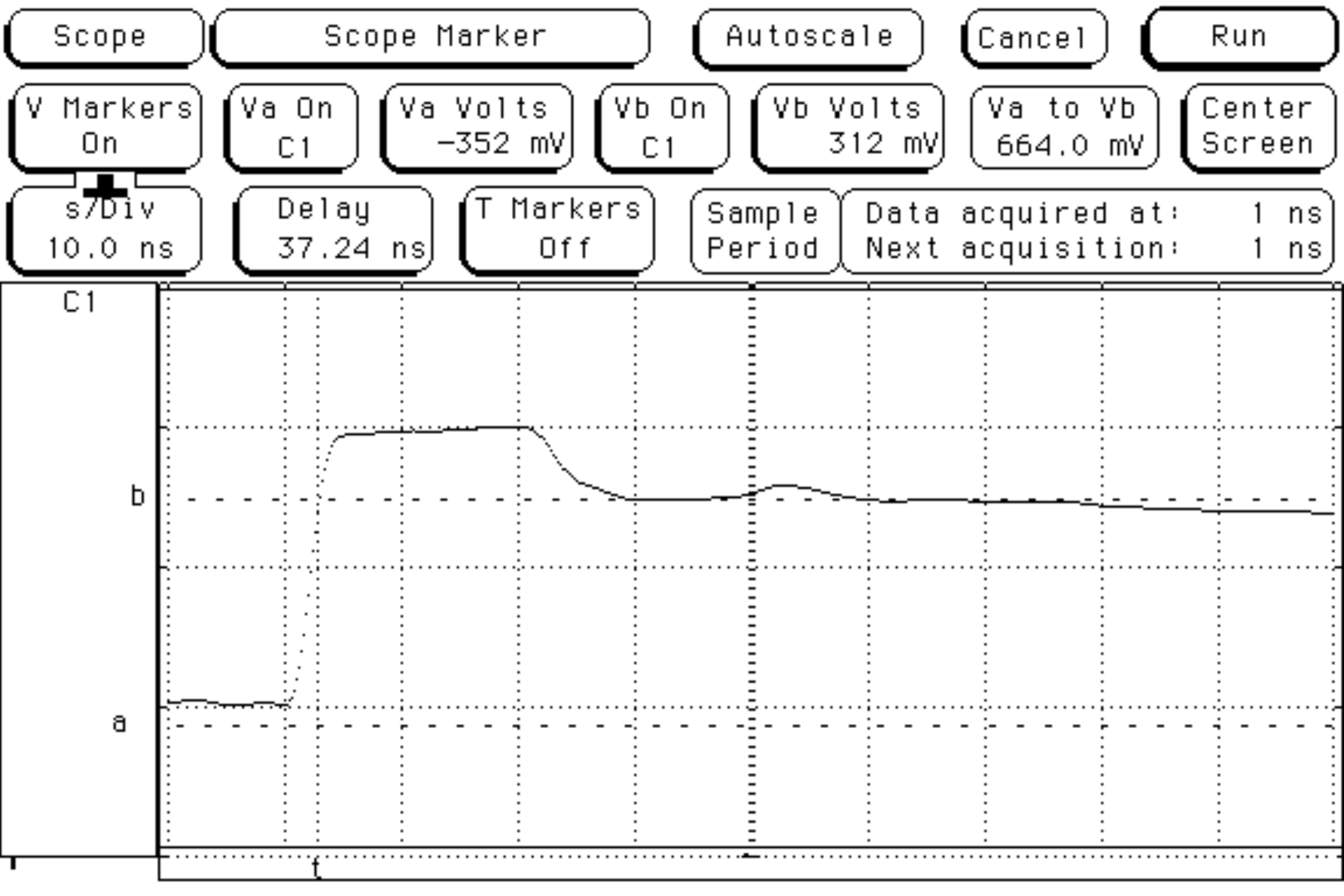}
\caption{Oscilloscope Output Using a 150cm cable with the TAP Attached}
\label{fig:oscillo}
\end{figure*}

A measurement was made with the TAP attached to the cable and a Raspberry Pi attached to the TAP, the result of this measurement is depicted in Figure~\ref{fig:oscillo} From this picture follows that no new reflections are introduced and no degradation to the signal is introduced other than the overall attenuation of the signal. This experiment shows that the TAP and attached Raspberry Pi do not introduce significant distortion to the signals present on the cable that TAP is attached to.
\section{Conclusion}
\label{conclusion}

From investigations conducted by NHTCU, demand emerged for a solution to the problem of identifying computer systems in a triage setting at data centres. In this paper, a device is presented that utilises passive interception of Ethernet network traffic to solve this problem. Prior work in the area was considered too invasive, i.e., where devices were placed in line with the connection that was to be intercepted. In this paper, it has been demonstrated that placing devices in line with the observed computer system's connection has the drawback that properties of the connection may change. A change in the connections properties may alert a suspect in the investigation. The passive method proposed in this method was evaluated for its impact to the reliability of the existing connection with good results.

\subsection{Future Work}
\label{futurework}

Future work is planned on addressing the problem with interception of Gigabit Ethernet. The electrical characteristics of Gigabit Ethernet seem  to suggest that a solution would require a device placed in line. Experiments with passive devices that may lead to methods of passive interception of Gigabit Ethernet links have left open for future work. The option to import case data from an USB memory device is desirable when plug-in modules require more complex data. The currently developed plug-in modules only require the input of data that is low in complexity and has a regular format that allows for fast and easy entry of this data using the two button user interface of the device.



%
%



\bibliographystyle{IEEEtran}
\bibliography{IEEEabrv,wiretapbibfile}
%



\end{document}